\documentclass[twocolumn,preprintnumbers,amsmath,amssymb]{revtex4}
\usepackage{graphicx}
\usepackage{epsfig}
\usepackage{amsmath}
\usepackage{amssymb}
\usepackage{dcolumn}

\begin{document}

\title{Representation of Quantum Field Theory by Elementary Quantum Information}

\author{Martin Kober}
\email{kober@fias.uni-frankfurt.de}

\affiliation{Frankfurt Institute for Advanced Studies (FIAS),
Johann Wolfgang Goethe-Universit\"at,
Ruth-Moufang-Strasse 1, 60438 Frankfurt am Main, Germany}

\date{\today}

\begin{abstract}
In this paper is considered relativistic quantum field theory expressed by elementary
units of quantum information as they are considered as fundamental entity of nature by Carl
Friedrich von Weizsaecker. Through quantization of a Weyl spinor describing an elementary
unit of quantum information and consisting of four real components one obtains four
pairs of creation and annihilation operators acting in a tensor space of states containing many
units of quantum information. There can be constructed position and momentum operators from the
creation and annihilation operators and based on these operators the Poincare group can be
represented in this abstract tensor space of quantum information. A general state in the tensor
space can be mapped to a state in Minkowski space-time by using the position representation
of the eigenstates of the occupation number operators which correspond to the eigenstates of
the harmonic oscillator. This yields a description of relativistic quantum mechanics.
Quantization of the coefficients of a general state in the tensor space leads to many
particle theory and thus to quantum field theory.
\end{abstract}

\maketitle

\section{Introduction}

In his program of a reconstruction of physics Carl Friedrich von Weizsaecker makes the attempt to reconstruct
physics from general quantum theory, which is considered as general theory of human knowledge about nature and
accordingly quantum theory is interpreted as consequence of general postulates about the structure of human knowledge \cite{Weizsaecker:1955},\cite{Weizsaecker:1971},\cite{Weizsaecker:1985},\cite{Weizsaecker:1992},\cite{Goernitz:1992},\cite{Lyre:1994eg},\cite{Lyre:1995gm},\cite{Lyre:2003tr}.
This means nothing else but that within this program physical objects and their properties shall be inferred from
abstract quantum information interpreted as fundamental entity of nature, which can be resolved into binary
alternatives called ur-alternatives by von Weizsaecker. This denotation arises from the fundamental role
the binary alternatives play in this approach, where is introduced a tensor space of many ur-alternatives, which are
elementary units of quantum information.
According to von Weizsaecker from the mathematical structure of the ur-alternatives, which can be described by
Weyl spinors, the existence of position space as representation of physical relations between objects consisting
of this elementary quantum information can be derived. This property is related to the fact that the symmetry
group of two dimensional complex Hilbert spaces is the special unitary group in two dimensions, the $SU(2)$, which is
isomorphic to the $SO(3)$, the rotation group in a three dimensional real vector space as it is realized in nature.
The group of general linear transformations in two complex dimensions, the $SL(2,\mathbb{C})$, is isomorphic to the Lorentz
group, which means that the structure of Minkowski space-time is also implicitly contained in a two dimensional
complex vector space \cite{Penrose:1985jw},\cite{Penrose:1986ca} describing the possible states of an
elementary unit of quantum information.
  
In the present paper is considered quantum field theory represented by elementary units of quantum information as they
are considered as fundamental entity of nature by von Weizsaecker. Concerning the basic concepts it is treated according
to von Weizsaecker. There is considered a tensor space describing many units of quantum information, but it is developed
a new method to obtain a representation of the states of the tensor space in usual space-time.
Since every physical state contains information and information can be represented by binary alternatives, it has to be possible to describe any quantum state including states of any particle or quantum field by elementary units of quantum information. If one wants to describe a physical object, a particle for example, one needs of course more than just
single units of quantum information. Therefore it is necessary to consider a Hilbert space of quantum states containing
many quantum information. Such a space is obtained by quantizing the Weyl spinor describing a single unit of quantum information. By quantizing the Weyl spinor the several components of the Weyl spinor describing the state of the unit of
quantum information become creation and annihilation operators. These creation and annihilation operators create and annihilate units of quantum information in the corresponding basis state of a single unit of quantum information the component refers to and therefore one is led to a Hilbert space, which is the tensor space of many units of quantum information and thus represents the tensor space referring to the vector space consisting of the possible states of single Weyl spinors. The creation and annihilation operators act in this Hilbert space and transform the basis states, which are the states with sharp occupation
numbers of units of quantum information in the several possible basis states of a single unit of quantum information, into
each other. The state of a single particle can be represented by a general state in such a Hilbert space of many units of
quantum information.

If quantum field theory shall be derived or at least be represented by quantum information, it is of course necessary
to relate states in the tensor space of elementary quantum information to states referring to space-time.
More exactly the states in the tensor space have to be represented in space-time.
The method to achieve a transition from tensor space to space-time as it is considered in this paper consists in the
definition of four pairs of position and momentum operators from the four pairs of creation and annihilation operators
obtained from the quantization of single units of quantum information.
As already mentioned a basis of states in the tensor space of elementary units of quantum information is given by all occupation numbers of the several basis states of an elementary unit of quantum information. These states are eigenstates
of the occupation number operators and thus correspond to the eigenstates of the more dimensional harmonic
oscillator in quantum mechanics, which can be represented in position space. Accordingly a general state
in the tensor space of elementary quantum information, which can be expressed by a superposition of eigenstates of the
occupation number operators, can be mapped to a state in space-time by representing the eigenstates of the
harmonic oscillator in space-time. Thus the states in the tensor space obtained by the quantization of the elementary
units of quantum information can indeed be represented as wave functions in space-time. With the position and momentum
operators defined by the creation and annihilation operators in the tensor space of elementary quantum information
can be constructed the generators of the Poincare group, which can thus be represented in the tensor space. Since
particles are described by irreducible representations of the Poincare group in quantum field theory \cite{Wigner:1939},\cite{Weinberg:1995mt}, the states in tensor space appearing as wave functions in space-time
can be considered as states of particles. According to usual field quantization the corresponding quantum field
theory referring to elementary quantum information is obtained by quantizing the coefficients of a general
state of a single particle, which corresponds to the wave function in usual quantum mechanics. This leads to
a theory of many particles and thus to quantum field theory. The basis states of the new Hilbert space are
defined by the occupation numbers referring to the basis states in the tensor space of elementary quantum information.

\section{Tensor Space of Elementary Quantum Information}

A single unit of quantum information is described by a Weyl spinor and thus by a normalized
two dimensional complex vector,

\begin{equation}
u=\left(\begin{matrix}u_1 \\ u_2 \end{matrix}\right)=\left(\begin{matrix} a+bi\\c+di \end{matrix}\right),
\end{equation}
which contains four real components denoted by $a$, $b$, $c$ and $d$, for which holds of course the following
relation as normalization condition: $a^2+b^2+c^2+d^2=1$.
Real physical objects naturally contain much quantum information and because of this there has to be considered
a Hilbert space of states containing much quantum information. To obtain such a Hilbert space of states
containing much quantum information and not just a single unit, it is necessary to quantize the elementary
unit of quantum information again, which means that it becomes an operator, $u \rightarrow \hat u$.
This quantization is performed as usual by postulating canonical commutation relations for the unit of quantum
information. Especially there is postulated the following commutation relation:

\begin{equation}
\left[\hat u_r,\hat u_s^{\dagger}\right]=\delta_{rs}\quad,\quad r,s=1,2,
\label{commutation_relation_Weyl}
\end{equation}
defining the quantum properties of the operator of the elementary unit of quantum information $\hat u$
and its hermitian adjoint $\hat u^{\dagger}$. This commutation relation ($\ref{commutation_relation_Weyl}$) can be
realized, if the real components of the corresponding Weyl spinor defined above, $a$, $b$, $c$
and $d$, become operators fulfilling the following commutation relations:

\begin{equation}
[\hat a,\hat a^{\dagger}]=1,\quad [\hat b,\hat b^{\dagger}]=1,\quad
[\hat c,\hat c^{\dagger}]=1,\quad [\hat d,\hat d^{\dagger}]=1.
\label{commutators_abcd}
\end{equation}
All other commutators between the operators $\hat a$, $\hat b$, $\hat c$ and $\hat d$ as well as the corresponding
hermitian adjoint operators $\hat a^{\dagger}$, $\hat b^{\dagger}$, $\hat c^{\dagger}$ and $\hat d^{\dagger}$ are
assumed to vanish. These operators accordingly behave as creation and annihilation operators creating and
annihilating elementary units of quantum information in the corresponding basis states of a single unit of quantum
information and thus relating different states within a tensor space of many units of quantum information.
The basis states within this tensor space are given by the number of units of quantum information, which are
in the four basis states of a single unit of quantum information, $|N_a,N_b,N_c,N_d \rangle$. 
Accordingly these states are related to each other as follows:

\begin{eqnarray}
\hat a|N_a,N_b,N_c,N_d \rangle &=& \sqrt{N_a}\ |N_a-1,N_b,N_c,N_d \rangle, \nonumber\\
\hat a^{\dagger}|N_a,N_b,N_c,N_d \rangle &=& \sqrt{N_a+1}\ |N_a+1,N_b,N_c,N_d \rangle, \nonumber\\
\nonumber\\
\hat b|N_a,N_b,N_c,N_d \rangle &=& \sqrt{N_b}\ |N_a,N_b-1,N_c,N_d \rangle, \nonumber\\
\hat b^{\dagger}|N_a,N_b,N_c,N_d \rangle &=& \sqrt{N_b+1}\ |N_a,N_b+1,N_c,N_d \rangle, \nonumber\\
\nonumber\\
\hat c|N_a,N_b,N_c,N_d \rangle &=& \sqrt{N_c}\ |N_a,N_b,N_c-1,N_d \rangle, \nonumber\\
\hat c^{\dagger}|N_a,N_b,N_c,N_d \rangle &=& \sqrt{N_c+1}\ |N_a,N_b,N_c+1,N_d \rangle, \nonumber\\
\nonumber\\
\hat d|N_a,N_b,N_c,N_d \rangle &=& \sqrt{N_d}\ |N_a,N_b,N_c,N_d-1 \rangle, \nonumber\\
\hat d^{\dagger}|N_a,N_b,N_c,N_d \rangle &=& \sqrt{N_d+1}\ |N_a,N_b,N_c,N_d+1 \rangle.\nonumber\\
\end{eqnarray}
They are created from a vacuum state $|0\rangle$ containing no quantum information
and being defined by the condition

\begin{equation}
\hat a|0\rangle=\hat b|0\rangle=\hat c|0\rangle=\hat d|0\rangle=0.
\end{equation}
One obtains an arbitrary basis state in the tensor space, $|N_a,N_b,N_c,N_d \rangle$, if the corresponding
creation operators $\hat a^{\dagger}$, $\hat b^{\dagger}$, $\hat c^{\dagger}$ and $\hat d^{\dagger}$ are
applied $N_a$, $N_b$, $N_c$ and $N_d$ times to the vacuum state $|0\rangle$,

\begin{equation}
|N_a,N_b,N_c,N_d \rangle=\frac{\hat a^{\dagger N_a} \hat b^{\dagger N_b}\hat c^{\dagger N_c} \hat d^{\dagger N_d}}
{\sqrt{N_a !}\sqrt{N_b !}\sqrt{N_c !}\sqrt{N_d !}}|0\rangle.
\end{equation}
The commutation relations of the creation and annihilation operators ($\ref{commutators_abcd}$) imply that the units
of quantum information obey Bose statistics. If they would obey Fermi statistics, there could according to the Pauli
exclusion principle only exist states containing not more than four units of quantum information corresponding to the
four possible basis states of one single unit. A basis state in the tensor space can be represented as the tensor
product of four independent states, $|N_a \rangle$, $|N_b \rangle$, $|N_c \rangle$ and $|N_d \rangle$, referring
to each pair of creation and annihilation operators. These states are eigenstates of the four occupation number
operators referring to the four basis states of a single unit of quantum information and thus fulfil the following
eigenvalue equations:

\begin{eqnarray}
\hat a^{\dagger} \hat a |N_a \rangle=N_a |N_a \rangle \quad,\quad
\hat b^{\dagger} \hat b |N_b \rangle=N_b |N_b \rangle,
\nonumber\\
\hat c^{\dagger} \hat c |N_c \rangle=N_c |N_c \rangle \quad,\quad 
\hat d^{\dagger} \hat d |N_d \rangle=N_d |N_d \rangle.       
\label{eigenstates_occupation_number_operators}
\end{eqnarray}
The inner product of these eigenstates with each other is defined according to

\begin{eqnarray}
\langle N_a^{\prime}|N_a \rangle&=&\delta_{N_a N_a^{\prime}}\quad,\quad
\langle N_b^{\prime}|N_b \rangle=\delta_{N_b N_b^{\prime}},\nonumber\\
\langle N_c^{\prime}|N_c \rangle&=&\delta_{N_c N_c^{\prime}}\quad,\quad
\langle N_d^{\prime}|N_d \rangle=\delta_{N_d N_d^{\prime}}.
\end{eqnarray}
Since a basis state in the complete tensor space can be expressed as the tensor product of the eigenstates
of the four occupation number operators defined in ($\ref{eigenstates_occupation_number_operators}$),

\begin{equation}
|N_a,N_b,N_c,N_d\rangle=|N_a \rangle \otimes |N_b \rangle \otimes |N_c \rangle \otimes |N_d \rangle,
\end{equation}
the inner product of two basis states in the complete tensor space can accordingly be expressed as 

\begin{eqnarray}
&&\langle N_a^{\prime},N_b^{\prime},N_c^{\prime},N_d^{\prime}|N_a,N_b,N_c,N_d\rangle\nonumber\\
&=&\langle N_a^{\prime}|N_a \rangle \langle N_b^{\prime}|N_b \rangle
\langle N_c^{\prime}|N_c \rangle \langle N_d^{\prime}|N_d \rangle\nonumber\\
&=&\delta_{N_a^{\prime} N_a}\delta_{N_b^{\prime} N_b}\delta_{N_c^{\prime} N_c}\delta_{N_d^{\prime} N_d}.
\label{inner_product_basis_states}
\end{eqnarray}
A general state $|\Psi \rangle$ in the Hilbert space of many units of quantum information corresponds
of course to an arbitrary superposition of the basis states,

\begin{equation}
|\Psi \rangle=\sum_{N_a,N_b,N_c,N_d} c(N_a,N_b,N_c,N_d)|N_a,N_b,N_c,N_d\rangle,
\label{general_state}
\end{equation}
where $c(N_a,N_b,N_c,N_d)$ denotes arbitrary coefficients, which can be described as
projections on the general state to the basis states,

\begin{equation}
c(N_a,N_b,N_c,N_d)=\langle N_a,N_b,N_c,N_d| \Psi \rangle.
\end{equation}
To be able to give a more clear arranged notation, it will be used
the following notation for a general state below:

\begin{equation}
|\Psi \rangle=\sum_N c(N)|N \rangle \quad,\quad N\ \widehat{=}\ (N_a,N_b,N_c,N_d).
\label{general_state_short}
\end{equation}
By using ($\ref{inner_product_basis_states}$), ($\ref{general_state}$) and ($\ref{general_state_short}$) the inner
product of two general states in the tensor space can be written as

\begin{eqnarray}
\langle \chi|\psi \rangle &=&\sum_{N^{\prime}}\sum_{N} d^{*}(N^{\prime})c(N)\langle N^{\prime}|N\rangle\nonumber\\
&=&\sum_{N^{\prime}}\sum_{N} d^{*}(N^{\prime})c(N)\delta_{N^{\prime} N}\nonumber\\
&=&\sum_{N} d^{*}(N)c(N).
\end{eqnarray}

\section{Construction of Position and Momentum Operators of Single Particles}

By introducing the new components $a_x$, $a_y$, $a_z$ and $a_t$ being defined by
the components $a$, $b$, $c$ and $d$ as follows:

\begin{eqnarray}
a_x=\frac{1}{2}\left(a-b+c-d\right),\quad
a_y=\frac{1}{2}\left(a-b-c+d\right),
\nonumber\\
a_z=\frac{1}{2}\left(a+b-c-d\right),\quad
a_t=\frac{1}{2}\left(a+b+c+d\right),
\end{eqnarray}
the Weyl spinor $u$ describing an elementary unit of quantum information can be expressed as

\begin{equation}
u=\left(\begin{matrix} \frac{1}{2}\left(a_x+a_y+a_z+a_t\right)
+\frac{1}{2}\left(-a_x-a_y+a_z+a_t\right)i\\
\frac{1}{2}\left(a_x-a_y-a_z+a_t\right)
+\frac{1}{2}\left(-a_x+a_y-a_z+a_t\right)i\end{matrix}\right).
\end{equation}
The new components $a_x$, $a_y$, $a_z$ and $a_t$ refer to linear combinations of eigenvectors
to the Pauli matrices

\begin{equation}
\sigma_x=\left(\begin{matrix}0&1\\1&0\end{matrix}\right),
\sigma_y=\left(\begin{matrix}0&-i\\i&0\end{matrix}\right),
\sigma_z=\left(\begin{matrix}1&0\\0&-1\end{matrix}\right),
\sigma_t=\left(\begin{matrix}1&0\\0&1\end{matrix}\right),
\end{equation}
in the Weyl spinor space of an elementary unit of quantum information.
After quantization the corresponding operators $\hat a_x$, $\hat a_y$, $\hat a_z$ and $\hat a_t$
fulfil the same commutation relations as the operators $\hat a$, $\hat b$, $\hat c$ and $\hat d$,

\begin{equation}
\left[\hat a_x,\hat a_x^{\dagger}\right]=1,\quad
\left[\hat a_y,\hat a_y^{\dagger}\right]=1,\quad
\left[\hat a_z,\hat a_z^{\dagger}\right]=1,\quad
\left[\hat a_t,\hat a_t^{\dagger}\right]=1,
\end{equation}
and therefore constitute the same tensor space referring to elementary quantum information, which has been considered in the last section, but now with respect to a new basis in the Weyl spinor space of a single elementary unit of quantum information.
The corresponding new basis states of the tensor space are accordingly defined by the occupation numbers
$N_x$, $N_y$, $N_z$ and $N_t$. The properties of the tensor space remain exactly the same as considered in the last section.  
From the annihilation operators $\hat a_x$, $\hat a_y$, $\hat a_z$ and $\hat a_t$ and the corresponding creation operators
$\hat a_x^{\dagger}$, $\hat a_y^{\dagger}$, $\hat a_z^{\dagger}$ and $\hat a_t^{\dagger}$ can now be defined hermitian
operators according to

\begin{eqnarray}
\hat x=\frac{1}{\sqrt{2}}\left(\hat a_x+\hat a_x^{\dagger}\right)\quad,\quad
\hat p_x=-\frac{i}{\sqrt{2}}\left(\hat a_x-\hat a_x^{\dagger}\right),\nonumber\\
\hat y=\frac{1}{\sqrt{2}}\left(\hat a_y+\hat a_y^{\dagger}\right)\quad,\quad
\hat p_y=-\frac{i}{\sqrt{2}}\left(\hat a_y-\hat a_y^{\dagger}\right),\nonumber\\
\hat z=\frac{1}{\sqrt{2}}\left(\hat a_z+\hat a_z^{\dagger}\right)\quad,\quad
\hat p_z=-\frac{i}{\sqrt{2}}\left(\hat a_z-\hat a_z^{\dagger}\right),\nonumber\\
\hat t=\frac{1}{\sqrt{2}}\left(\hat a_t+\hat a_t^{\dagger}\right)\quad,\quad
\hat p_t=-\frac{i}{\sqrt{2}}\left(\hat a_t-\hat a_t^{\dagger}\right),
\label{position_momentum_operator}
\end{eqnarray}
fulfilling a Heisenberg algebra like position and momentum operators,

\begin{equation}
[\hat x,\hat p_x]=i,\quad [\hat y,\hat p_y]=i,\quad [\hat z,\hat p_z]=i,\quad [\hat t,\hat p_t]=i,
\end{equation}
which are therefore isomorphic to such operators. It is now a postulate to identify these operators with the real
position and momentum operators referring to Minkowski space-time. If one wants conversely to express the creation
and annihilation operators by the position and momentum operators, one obtains the following equations:

\begin{eqnarray}
\hat a_x=\frac{1}{\sqrt{2}}\left(\hat x+i\hat p_x\right)\quad,\quad
\hat a_x^{\dagger}=\frac{1}{\sqrt{2}}\left(\hat x-i\hat p_x\right),\nonumber\\
\hat a_y=\frac{1}{\sqrt{2}}\left(\hat y+i\hat p_y\right)\quad,\quad
\hat a_y^{\dagger}=\frac{1}{\sqrt{2}}\left(\hat y-i\hat p_y\right),\nonumber\\
\hat a_z=\frac{1}{\sqrt{2}}\left(\hat z+i\hat p_z\right)\quad,\quad
\hat a_z^{\dagger}=\frac{1}{\sqrt{2}}\left(\hat z-i\hat p_z\right),\nonumber\\
\hat a_t=\frac{1}{\sqrt{2}}\left(\hat t+i\hat p_t\right)\quad,\quad
\hat a_t^{\dagger}=\frac{1}{\sqrt{2}}\left(\hat t-i\hat p_t\right).
\label{creation_annihilation_position_momentum_operator}
\end{eqnarray}
Of course the position and momentum operators constructed by the creation and annihilation operators referring to
the abstract tensor space of elementary quantum information ($\ref{position_momentum_operator}$) lead to Minkowski
space-time, since they can naturally be represented in Minkowski space-time. By referring to this representation
in Minkowski space-time, the creation and annihilation operators look as follows:

\begin{eqnarray}
\hat a_x=\frac{1}{\sqrt{2}}\left(x+\frac{\partial}{\partial x}\right)\quad,\quad
\hat a_x^{\dagger}=\frac{1}{\sqrt{2}}\left(x-\frac{\partial}{\partial x}\right),\nonumber\\
\hat a_y=\frac{1}{\sqrt{2}}\left(y+\frac{\partial}{\partial y}\right)\quad,\quad
\hat a_y^{\dagger}=\frac{1}{\sqrt{2}}\left(y-\frac{\partial}{\partial y}\right),\nonumber\\
\hat a_z=\frac{1}{\sqrt{2}}\left(z+\frac{\partial}{\partial z}\right)\quad,\quad
\hat a_z^{\dagger}=\frac{1}{\sqrt{2}}\left(z-\frac{\partial}{\partial z}\right),\nonumber\\
\hat a_t=\frac{1}{\sqrt{2}}\left(t+\frac{\partial}{\partial t}\right)\quad,\quad
\hat a_t^{\dagger}=\frac{1}{\sqrt{2}}\left(t-\frac{\partial}{\partial t}\right).
\end{eqnarray}

\section{Transition from States in Tensor space to States in Space-Time}

In the last section it has been shown that it is possible to map the four pairs of creation and annihilation
operators acting in the tensor space referring to a Weyl spinor as basis space to position and momentum operators
referring to a Minkowski space-time. This means that the tensor space of elementary quantum information is
isomorphic to Minkowski space-time and therefore it has to be possible to represent the states in the tensor space
as wave functions on Minkowski space-time.
To obtain the representation in Minkowski space-time there has to be used the mathematical description of the more
dimensional quantum theoretical harmonic oscillator. The harmonic oscillator can be described by introduction
of creation and annihilation operators, which are related to the position and momentum operators according to the
consideration of this paper ($\ref{position_momentum_operator}$),
($\ref{creation_annihilation_position_momentum_operator}$).
The Hamilton operator of the harmonic oscillator corresponds then to the occupation number operator and an additional
vacuum expectation value and the eigenstates of the Hamilton operator can be represented in position space by using the
hermite polynoms. In the scenario of this paper is defined no Hamilton operator, but the eigenstates of the Hamilton
operator of the harmonic oscillator correspond to the basis states in the tensor space, which can therefore be
represented in position space, in Minkowski space-time in this case to be more specific.
One has to begin with the separated eigenstates of the occupation number operators,
which can be represented as in case of the harmonic oscillator in position space according to

\begin{eqnarray}
\varphi_{N_x}(x)&=&\langle x|N_x \rangle=\frac{1}{\left(N_x! 2^{N_x} \pi^2\right)}
\left(x-\frac{\partial}{\partial x}\right)^{N_x} \exp\left(-\frac{x^2}{2}\right), \nonumber\\
\varphi_{N_y}(y)&=&\langle y|N_y \rangle=\frac{1}{\left(N_y! 2^{N_y} \pi^2\right)}
\left(y-\frac{\partial}{\partial y}\right)^{N_y} \exp\left(-\frac{y^2}{2}\right), \nonumber\\
\varphi_{N_z}(z)&=&\langle z|N_z \rangle=\frac{1}{\left(N_z! 2^{N_z} \pi^2\right)}
\left(z-\frac{\partial}{\partial z}\right)^{N_z} \exp\left(-\frac{z^2}{2}\right), \nonumber\\
\varphi_{N_t}(t)&=&\langle t|N_t \rangle=\frac{1}{\left(N_t! 2^{N_t} \pi^2\right)}
\left(t-\frac{\partial}{\partial t}\right)^{N_t} \exp\left(-\frac{t^2}{2}\right).
\end{eqnarray}
An arbitrary state in the complete tensor space can then accordingly be represented as a superposition of
the tensor product of these projections to the eigenstates of the four position operators

\begin{eqnarray}
&&\Psi(x,y,z,t)=\langle x,y,z,t|\Psi\rangle \nonumber\\
&=&\sum_{N}c(N_x,N_y,N_z,N_t)
\langle x|N_x \rangle \langle y|N_y \rangle \langle z|N_z \rangle \langle t|N_t \rangle\nonumber\\
&=&\sum_{N}c(N_x,N_y,N_z,N_t) \varphi_{N_x}(x)\varphi_{N_y}(y)\varphi_{N_z}(z)\varphi_{N_t}(t),
\nonumber\\
\end{eqnarray}
where it has been defined: $|x,y,z,t \rangle=|x\rangle \otimes |y\rangle \otimes |z\rangle \otimes |t\rangle$.
This means that a general state in the tensor space ($\ref{general_state}$) is isomorphic to a wave function
in Minkowski space-time. Below a general state in the tensor space represented as a wave function in Minkowski
space-time will be written as following: 

\begin{equation}
\Psi(X)=\Psi(x,y,z,t)\quad,\quad X=(x,y,z,t),
\label{coordinate}
\end{equation}
and the product of the eigenstates of the four occupation number operators will be 
expressed as

\begin{equation}
\varphi_N(X)=\varphi_{N_x}(x)\varphi_{N_y}(y)\varphi_{N_z}(z)\varphi_{N_t}(t).
\label{basis_state_wave_function}
\end{equation}

\section{Representation of the Poincare Group in Tensor Space}

If a general state in the tensor space can be represented in Minkowski space-time, it has to be possible
to represent the Poincare group in the tensor space. In this section there shall be given the representation
of the Poincare group in the tensor space. A transformation in the tensor space yields then a corresponding
transformation concerning the representation of the state in Minkowski space-time. 
The Poincare group consists of the four space-time translations, the rotations in the three-dimensional
spatial subspace of Minkowski space-time and the Lorentz boosts referring to the the three spatial directions.
Thus it is described by ten generators. The four generators of space-time translations are the momentum
operators and the energy operator given in ($\ref{position_momentum_operator}$),

\begin{eqnarray}
P_0=\hat p_t=-\frac{i}{\sqrt{2}}\left(\hat a_t-\hat a_t^{\dagger}\right),\nonumber\\
P_1=\hat p_x=-\frac{i}{\sqrt{2}}\left(\hat a_x-\hat a_x^{\dagger}\right),\nonumber\\
P_2=\hat p_y=-\frac{i}{\sqrt{2}}\left(\hat a_y-\hat a_y^{\dagger}\right),\nonumber\\
P_3=\hat p_z=-\frac{i}{\sqrt{2}}\left(\hat a_z-\hat a_z^{\dagger}\right).
\end{eqnarray}
From these operators there can be constructed the generators of the Lorentz group consisting
of the three angular momentum operators generating the three-dimensional rotation group and
the three Lorentz boost generators,

\begin{eqnarray}
M_{23}=\hat y \hat p_z-\hat z \hat p_y=i\left(\hat a_y^{\dagger} \hat a_z-\hat a_z^{\dagger} \hat a_y\right),\nonumber\\
M_{13}=\hat z \hat p_x-\hat x \hat p_z=i\left(\hat a_z^{\dagger} \hat a_x-\hat a_x^{\dagger} \hat a_z\right),\nonumber\\
M_{12}=\hat x \hat p_y-\hat y \hat p_x=i\left(\hat a_x^{\dagger} \hat a_y-\hat a_y^{\dagger} \hat a_x\right),\nonumber\\
M_{01}=\hat t \hat p_x+\hat x \hat p_t=i\left(\hat a_t \hat a_x-\hat a_x^{\dagger} \hat a_t^{\dagger}\right),\nonumber\\
M_{02}=\hat t \hat p_y+\hat y \hat p_t=i\left(\hat a_t \hat a_y-\hat a_y^{\dagger} \hat a_t^{\dagger}\right),\nonumber\\
M_{03}=\hat t \hat p_z+\hat z \hat p_t=i\left(\hat a_t \hat a_z-\hat a_z^{\dagger} \hat a_t^{\dagger}\right).
\end{eqnarray}
The generators of the Poincare group fulfil the following Lie Algebra defining its characteristic group structure,

\begin{eqnarray}
\left[P_\mu,P_\nu \right]&=&0,\quad\quad\quad\quad\quad\quad\quad\quad\quad\quad\quad \mu,\nu,\rho,\sigma=0...3,\nonumber\\
\left[M_{\mu\nu},P_\rho\right]&=&i\eta_{\mu\rho} P_\nu-i\eta_{\nu\rho} P_\mu,\nonumber\\
\left[M_{\mu\nu},M_{\rho\sigma}\right]&=&-i\eta_{\mu\rho}M_{\nu\sigma}+i\eta_{\mu\sigma}M_{\nu\rho}
-i\eta_{\nu\rho}M_{\mu\sigma}+i\eta_{\nu\sigma}M_{\mu\rho}.\nonumber\\
\end{eqnarray}
This means that a general Poincare transformation applied to a general quantum state in the
tensor space is of the following shape:

\begin{eqnarray}
|\Psi \rangle\ \rightarrow\ &&\exp\left[
ia_\mu P^\mu(\hat a,\hat a^{\dagger})+i\omega_{\mu\nu}M^{\mu\nu}(\hat a,\hat a^{\dagger})\right]|\Psi \rangle \nonumber\\
&&=\sum_{N} c(N)\exp\left[i a_\mu P^\mu(\hat a,\hat a^{\dagger})
+i\omega_{\mu\nu}M^{\mu\nu}(\hat a,\hat a^{\dagger})\right]
|N\rangle.\nonumber\\
\end{eqnarray}

\section{Quantum Field Theory}

In the last section an arbitrary state in the tensor space of elementary quantum information has been mapped to
four-dimensional space-time where can be represented the Lorentz group and thus it has been mapped
to Minkowski space-time. The resulting state can be considered as a wave function describing a particle.
A basis state in the tensor space is defined by the occupation number with respect to the four basis
states of a single unit of quantum information.
To obtain a description of quantum field theory, which is based on the elementary quantum information,
the wave function as representation of
a general state in the tensor space has to become an operator by itself, which acts on states in a new
Hilbert space. The wave function obtained as representation of a state in the Hilbert space of many units
of elementary quantum information in the consideration developed here is transformed to an operator by
postulating commutation relations of the coefficients defining a general state in tensor space as
superposition of the corresponding basis states,

\begin{equation}
\left[\hat c(N),\hat c^{\dagger}(N^{\prime})\right]
=i\delta_{N N^{\prime}}.
\end{equation}
Thus the coefficients become operators $\hat c(N)$, which behave as annihilation operators in a new Hilbert space.
The corresponding hermitian adjoint operators $\hat c^{\dagger}(N)$ behave accordingly as creation operators.
The wave function in space-time acts after quantization as an operator in this Hilbert space,

\begin{equation}
\hat \Psi(x,y,z,t)=\sum_{N}\hat c(N)\ \varphi_{N_x}(x)\varphi_{N_y}(y)\varphi_{N_z}(z)\varphi_{N_t}(t).
\end{equation}
By using ($\ref{coordinate}$) it can be represented as

\begin{equation}
\hat \Psi(X)=\sum_{N}\hat c(N)\ \varphi_{N}(X).
\end{equation}
The corresponding vacuum state, $|0\rangle_{\mathcal{N}}$, is defined by

\begin{equation}
\hat c(N)|0\rangle_{\mathcal{N}}=0,\ \forall\ N.
\end{equation} 
The basis states of the Hilbert space of quantum field theory, $|\mathcal{N}\rangle$,
defined by the number of particles in the corresponding basis states of the tensor space, which is
according to the considerations above defined by a distribution of quantum information in the several
states of a single unit of quantum information, can be obtained by applying the corresponding
creation operators for particles, $\hat c(N)^{\dagger}$, $\mathcal{N}(N)$ times to the vacuum state
$|0\rangle_{\mathcal{N}}$,

\begin{equation}
|\mathcal{N}\rangle=\sum_{N}{\hat c^{\dagger}(N)}^{\mathcal{N}(N)}|0\rangle_{\mathcal{N}}.
\label{basis_quantum_field_theory}
\end{equation}
The corresponding scalar product between these basis states is accordingly defined as

\begin{equation}
\langle \mathcal{N}^{\prime}|\mathcal{N}\rangle
=\prod_{N}\delta_{\mathcal{N}^{\prime}(N)\mathcal{N}(N)}=\delta_{\mathcal{N}^{\prime} \mathcal{N}}.
\end{equation}
A general state in the Hilbert space of many particle theory or quantum field theory is defined
by a general superposition of basis states ($\ref{basis_quantum_field_theory}$) in the following way:

\begin{equation}
|\Phi\rangle=\sum_{\mathcal{N}}\mathcal{C}(\mathcal{N})|\mathcal{N}\rangle.
\end{equation}
It is also possible to formulate a propagator within the tensor space,
$\Delta(X,X^{\prime})$, which is defined as

\begin{equation}
\Delta(X,X^{\prime})=\langle 0|\hat \Psi(X)\hat \Psi^{\dagger}(X^{\prime})|0 \rangle.
\end{equation}
This expression for the propagator can be transformed by the following calculation:

\begin{eqnarray}
\Delta(X,X^{\prime})&=&\langle 0|\hat \Psi(X)\hat \Psi^{\dagger}(X^{\prime})|0 \rangle\nonumber\\
&=&\sum_{N,N^{\prime}}\varphi_N (X)\varphi^{*}_{N^{\prime}}(X^{\prime})
\langle 0|\hat c(N)\hat c^{\dagger}(N^{\prime})|0 \rangle \nonumber\\
&=&\sum_{N,N^{\prime}}\varphi_N (X)\varphi^{*}_{N^{\prime}}(X^{\prime})
\langle \Phi_{\mathcal{N}(N)=1}|\Phi^{\prime}_{\mathcal{N}(N^{\prime})=1}\rangle \nonumber\\
&=&\sum_{N,N^{\prime}}\varphi_N (X)\varphi^{*}_{N^{\prime}}(X^{\prime})
\delta_{N,N^{\prime}} \nonumber\\
&=&\sum_{N}\varphi_N (X)\varphi^{*}_{N}(X^{\prime}),
\label{calculation_propagator}
\end{eqnarray}
where $\varphi_N (X)$ describes according to ($\ref{basis_state_wave_function}$) the representation of a basis state
in the tensor space as wave function in space-time. In the calculation ($\ref{calculation_propagator}$) has been
used the following notation for a state of a single particle expressed in the Hilbert space of quantum field theory:

\begin{equation}
|\Phi_{\mathcal{N}(M)=1}\rangle\ \widehat{=}\ \begin{cases}\mathcal{N}(N)=1,\ \rm{if}\ N=M \\ \mathcal{N}(N)=0\ \rm{else}\end{cases}.
\end{equation}

\section{Conceptual Issues}

Both theories of contemporary fundamental physics, quantum theory as well as general relativity, suggest
that physical reality is nonlocal on a fundamental level. In general relativity the mathematical property
of diffeomorphism invariance is connected to the relationalistic attitude with respect to space-time, which
is suggested by the theory. In quantum theory the mathematical formalism and many phenomena
unambiguously reflect such a nonlocal property of nature. The postulates of general
quantum theory within the mathematical formulation given by Paul Adrien Maurice Dirac and Johann von Neumann
\cite{Dirac},\cite{Neumann} do not presuppose position space. This is the reason why it was possible to deal
with a quantum theory of abstract quantum information without presupposing space-time in this paper, but to
represent the obtained states in a mathematical space being isomorphic to real space-time and to postulate
that this representation space corresponds to real space-time. 
Phenomena like the Einstein Podolsky Rosen paradoxon or the double slit experiment show with respect to
concrete experiments that physical reality as it is described by quantum theory is nonlocal. In a state, where
two particles moving into two opponent directions are correlated with each other, the measurement of an observable
of one particle instantaneously influences the complete state and thus also the other particle. In the double
slit experiment with single particles the question through which slit the particle has moved already leads
to contradictions with the experiment and thus the concept of a trajectory of a particle is not appropriate
concerning quantum theoretical phenomena.
If these phenomena would be interpreted as phenomena, which presuppose the local causal structure of space-time,
they would explicitly contradict special relativity, since they contain an instantaneous exchange of physical
information. Therefore it seems to be necessary to assume that they refer to a level of reality where this structure
must not to be presupposed yet, but that this more fundamental quantum theoretical reality constitutes the macroscopic
local causal structure of space-time, which thus has to be derived from quantum theory. 
Since the mathematical formalism of general quantum theory seems to describe physics in the microcosm as far as
it is known correctly, it is not astonishing that the nonlocal character of quantum theory is contained in the
mathematical formalism as well as the concrete phenomena. The one is just the general theoretical representation
of the other.
If space-time is not considered as fundamental, then particles and fields as they are presupposed in particle physics
have to be replaced by a more fundamental entity. This entity could be the elementary quantum information, which is
constitutive within Carl Friedrich von Weizsaeckers idea of a reconstruction of physics containing the concept of
alternatives as general shape of human knowledge about nature.
The description of physical states by using elementary quantum information does not presuppose space-time, but
leads to a description of physical reality implicitly containing the structure of real space-time. Accordingly space-time
is considered as a kind of representation of a more fundamental reality of quantum theoretical relations, which can be
expressed by quantum information resolved into elementary units of quantum information. The conceptual issues concerning
this question of the nature of space-time are more elaborately treated in
\cite{Weizsaecker:1955}\cite{Weizsaecker:1971},\cite{Weizsaecker:1985},\cite{Weizsaecker:1992},\cite{Kober:2009},\cite{Kober:2010}.

\section{Summary and Discussion}

In this paper has been considered quantum field theory within a representation by elementary quantum information.
There has been defined a tensor space of quantum information as it appears within the quantum theory of ur-alternatives
of Carl Friedrich von Weizsaecker, which postulates abstract quantum information as fundamental
entity of nature. From the components of the quantized units of quantum information described by Weyl spinors, which become
operators annihilating units of quantum information in the corresponding basis state and thus constituting the tensor space,
have been defined position and momentum operators. It has been shown that by referring to these operators and using the eigenstates of the harmonic oscillator represented in position space it is possible to perform a transition from a general state in the tensor space of quantum information to position space.
This means that a general state in the tensor space can be represented as a wave function in Minkowski space-time describing
the state of a single particle. Another quantization of the amplitudes of the states in the tensor space leads to many particle
theory corresponding to quantum field theory. The theory gives the possibility to obtain a description of quantum field theory
related to a more fundamental purely quantum theoretical and thus nonlocal reality, which constitutes physical objects
appearing in space-time. Since the tensor space of elementary quantum information is discrete, it could become possible to
avoid the appearance of divergences from the beginning.\\


\begin{thebibliography}{99}

\bibitem{Weizsaecker:1955}
  C.~F.~von~Weizsaecker,
  ``Komplementarität und Logik,''
  Die Naturwissenschaften {\bf 42}, 521-529, 545-555 (1955).\\
  C.~F.~von~Weizsaecker,
  ``Komplementarität und Logik, II. Die Quantentheorie der einfachen Alternative,''
  Zeitschrift für Naturforschung {\bf 13a}, 245-253 (1958).\\
  C.~F.~von~Weizsaecker, E.~Scheibe und G.~Suessmann,
  ``Komplementarität und Logik, III. Mehrfache Quantelung,''
  Zeitschrift für Naturforschung {\bf 13a}, 705 (1958).

\bibitem{Weizsaecker:1971}
  C.~F.~von Weizsaecker, {\it Die Einheit der Natur},
  {\it Carl Hanser Verlag, Muenchen, 1971.}

\bibitem{Weizsaecker:1985}
  C.~F.~von Weizsaecker, {\it Aufbau der Physik},
  {\it Carl Hanser Verlag, Muenchen, 1985.}

\bibitem{Weizsaecker:1992}
  C.~F.~von Weizsaecker, {\it Zeit und Wissen},
  {\it Carl Hanser Verlag, Muenchen, 1992.}

\bibitem{Goernitz:1992}
  T.~Goernitz, D.~Graudenz und C.~F.~von~Weizsaecker,
  ``Quantum Field Theory of Binary Alternatives,''
  Int.\ J.\ Theor.\ Phys.\ {\bf 31}, 1929-1959 (1992). 

\bibitem{Lyre:1994eg}
 H.~Lyre,
 %``The Quantum Theory of Ur-Objects as a Theory of Information,''
 Int.\ J.\ Theor.\ Phys.\  {\bf 34} (1995) 1541
 [arXiv:quant-ph/9611048].
 %%CITATION = IJTPB,34,1541;%%

\bibitem{Lyre:1995gm}
 H.~Lyre,
 %``Multiple Quantization and the Concept of Information,''
 Int.\ J.\ Theor.\ Phys.\  {\bf 35} (1996) 2219
 [arXiv:quant-ph/9702047].
 %%CITATION = IJTPB,35,2219;%%

\bibitem{Lyre:2003tr}
  H.~Lyre,
  %``C.F. von Weizsaecker's reconstruction of physics: Yesterday, today, tomorrow,''
  %\href{http://www.slac.stanford.edu/spires/find/hep/www?irn=7014163}{SPIRES entry}
  {\it  In *Castell, L. (ed.) et al.: Time, quantum and information* 373-383.}

\bibitem{Penrose:1985jw}
  R.~Penrose and W.~Rindler,
  {\it Spinors and space-time. Vol. 1: Two spinor calculus and relativistic fields},
  %\href{http://www.slac.stanford.edu/spires/find/hep/www?irn=1396293}{SPIRES entry}
  Cambridge University Press, Cambridge 1984, p. 458.

\bibitem{Penrose:1986ca}
  R.~Penrose and W.~Rindler,
  {\it Spinors and space-time. Vol. 2: Spinor and twistor methods in space-time},
  %GEOMETRY,''
  %\href{http://www.slac.stanford.edu/spires/find/hep/www?irn=1653610}{SPIRES entry}
  Cambridge University Press, Cambridge 1986, p. 501.

\bibitem{Wigner:1939}
  E.~P.~Wigner,
  % On unitary representations of the inhomogeneous Lorentz group,
  Annals of Mathematics 40 (1939), 149-204

\bibitem{Weinberg:1995mt}
  S.~Weinberg, {\it The Quantum theory of fields. Vol. 1: Foundations},
  %\href{http://www.slac.stanford.edu/spires/find/hep/www?irn=3355144}{SPIRES entry}
  Cambridge University Press, Cambridge 1995, p. 609.

\bibitem{Dirac}
  P.A.M.~Dirac,
  {\it The principles of quantum mechanics}, Oxford University Press, Oxford 1958.

\bibitem{Neumann}
  J.~von~Neumann,
  {\it Mathematische Grundlagen der Quantenmechanik}, Berlin 1932.

\bibitem{Kober:2009}
  M.~Kober,
  ``Copenhagen Interpretation of Quantum Theory and the Measurement Problem,''
  [arXiv:0905.0408].

\bibitem{Kober:2010}
  M.~Kober,
  {\it Die Konstituierung der Raum-Zeit in einer einheitlichen Naturtheorie. Ueber die Beziehung
  der begrifflichen Grundlagen der Quantentheorie und der Allgemeinen Relativitaetstheorie},
  Suedwestdeutscher Verlag fuer Hochschulschriften, Saarbruecken 2011.

\end{thebibliography}
\end{document}